\documentclass[10pt]{article}

\usepackage{epsfig}

\def\lsim{\mathrel{\rlap{\lower4pt\hbox{\hskip1pt$\sim$}}
    \raise1pt\hbox{$<$}}}         
\def\gsim{\mathrel{\rlap{\lower4pt\hbox{\hskip1pt$\sim$}}
    \raise1pt\hbox{$>$}}}         

\def\overleftrightarrow#1{\vbox{\ialign{##\crcr
    $\leftrightarrow$\crcr
    \noalign{\kern 1pt\nointerlineskip}
    $\hfil\displaystyle{#1}\hfil$\crcr}}}

\def \lta {\mathrel{\vcenter {\hbox{$<$}\nointerlineskip\hbox{$\sim$}}}}

\newcommand{\pp}{\\[0.5cm]}
\newcommand{\be}{\begin{equation}}
\newcommand{\ee}{\end{equation}}
\newcommand{\bea}{\begin{eqnarray}}
\newcommand{\eea}{\end{eqnarray}}

\newcommand{\vk}{k}

\begin{document}

\hfill {\bf TUM/T39-03-03} \\
\vspace{0.01in}
\hfill {\bf ECT*-03-07} \\
\vspace{0.01in}
\hfill {March 2003} \\

\vspace{0.25in}

\begin{center}
{\bf \Large The QCD Running Coupling at Finite  \\ Temperature and Density\footnote{Work
supported in part by BMBF, GSI and by the European Commission under contract HPMT-CT-2001-00370.}}
\end{center}
\vspace{0.25 in}
\begin{center}
{R.A. Schneider}\\
{\small \em  Physik-Department, Technische Universit\"{a}t M\"{u}nchen, 85747 Garching,
Germany\\}
{\small \em  ECT*, Villa Tambosi, 38050 Villazzano (Trento), Italy\\}
%

%
\end{center}
\vspace{0.25 in}

\begin{abstract}
We present for the first time a self-contained calculation of the QCD running coupling at finite temperature and quark chemical potential, $\alpha_s(T, \mu)$, based on a semiclassical background field method. The hard thermal/dense loop  results on the Debye screening mass are recovered in a first approximation. The final result can be interpreted as the ordinary zero temperature running coupling, with momenta replaced by in-medium scales $\Lambda_*$: at high density and zero temperature, the quark scale is set by $\Lambda_* \approx 24.4 \ \mu$. At high temperature and moderate densities, the quark in-medium scale reads $\Lambda_*^2 \approx [2.91 \ T]^2 + [1.91 \ \mu]^2$, which resembles the naive phenomenological estimate $\Lambda_*^2 = (\pi T)^2 + \mu^2$.
\end{abstract} 
\vspace{0.25in}
\section{Introduction}
First steps in the QCD phase diagram have been made possible recently by progress in lattice calculations at finite temperature $T$ and baryochemical potential $\mu_B = 3\mu$, where $\mu$ is the quark chemical potential \cite{Fodor:2002a,Fodor:2002b,Allton:2002}. Results on the equation of state of QCD are now available for small to moderate chemical potentials $\mu \lta T_c$. Whereas the temperature region close to the phase transition is inherently non-perturbative and at the moment only treated within phenomenological models, both at $\mu = 0$ \cite{Peshier:1996,Levai:1998,Schneider:2001} and $\mu \neq 0$ \cite{Peshier:2002,Letessier:2003}, various perturbative resummation techniques apparently work quite well above roughly $3 \ T_c$ \cite{Andersen:2001,Blaizot:2001} and allow an interpretation of the thermodynamics in terms of weakly interacting quasiparticle partons that carry the same quantum numbers as the fundamental quarks and gluons. An extension of these calculations to finite chemical potential is straightforward and will allow further tests of the models by comparing their predictions to the mentioned lattice data.
\pp
However, since there is no scale around other than the renormalisation scale $\Lambda$ in the chiral limit of QCD, all the resummations basically do is to fix the magnitude of the partition function. The variation with respect to temperature is solely caused by the temperature-dependent running coupling constant; the normalised, dimensionless pressure, e.g., reads $p(T)/T^4 = f[\alpha_s(T/\Lambda)]$. In all perturbative calculations, $\alpha_s$ is assumed to take on the form of the common zero-temperature running coupling $\alpha_s(k^2)$, derived from one- or two-loop renormalisation group equations. The momentum scale $k$ is then usually chosen to resemble a typical thermal scale $\Lambda_*$, e.g. the lowest non-vanishing gluon Matsubara frequency $2 \pi T$. In \cite{Schneider:2002,Schneider:2003}, we have derived that this is indeed a sensible choice for long wavelengths provided the different Matsubara frequencies for bosons and fermions are taken into account: 
\be
\alpha_{s, \rm eff}(T, \Lambda) = \frac{\alpha_s(\Lambda)}{1 + {\displaystyle \frac{\alpha_s(\Lambda)}{12\pi} \left[ 11 N_c \log \left(\frac{[\Lambda_*^g]^2}{\Lambda^2} \right) - 2 N_f  \log \left(\frac{[\Lambda_*^f]^2}{\Lambda^2} \right) \right] }}, \label{alpha_s_T}
\ee
with 
\be
\Lambda_*^g =  \bar{\mathcal{A}}_g (2 \pi T), \quad \mbox{ where } \bar{\mathcal{A}}_g = \exp(- \gamma - 1/2 \log(3/8) + 1/44) \approx 0.938, \label{Lambda_g}
\ee
and
\be
\Lambda_*^f = \bar{\mathcal{A}}_f (\pi T), \quad \mbox{ where } \bar{\mathcal{A}}_f = \exp(- \gamma + 1/2) \approx 0.926 \label{Lambda_f}
\ee
(stars $_*$ denote in-medium quantities). The separation of thermal in-medium scales for the quarks and gluons arises naturally in our calculation and is also expected phenomenologically on second thought, but has not been taken into account in actual perturbative calculations so far. 
\pp
In the presence of a quark chemical potential, it is not so easy to argue what the relevant scale might be. First, one expects that $\mu$ modifies, to one-loop order, only the quark part of the running coupling, hence $\Lambda_*^f = f(T, \mu)$. Second, the fermionic Matsubara frequency is not an intuitive guide line anymore since it becomes complex: $\omega_n(T, \mu) = (2n+1)\pi T - i\mu$. From $| \omega_0 |^2$ or, equivalently, identity (\ref{FD_identity}), one might argue that 
\be
(\Lambda^f_*)^2 = (\pi T)^2 + \mu^2 \label{scale_1}
\ee
constitutes an appropriate in-medium scale for fermions. For $\mu \lta T$, the temperature would hence dominate the magnitude of $\Lambda_*^f$. On the other hand, one may define an average thermal momentum by
\be
\langle k \rangle_*  =  \frac{\displaystyle \int d^3 k |k| f_D(k - \mu)}{\displaystyle \int d^3 k f_D(k)} \simeq 3.15 \ T + 3 \mu + \mathcal{O}(\mu^2) \nonumber
\ee
for small $\mu/T$, and $f_D(k) = [\exp(k/T) + 1]^{-1}$. This result implies that $T$ and $\mu$ are on an equal footing, in contrast to our first estimate above. 
\pp
Also note that the in-medium momentum $\langle k \rangle_*$ for quarks will -- on general grounds -- always increase with the chemical potential. A larger $\langle k \rangle_*$ (or, equivalently, $\Lambda_*^f$), however, implies that the antiscreening property arising from the gluons will be even more compensated by the (ordinary) screening from the quarks than at $\mu =0$, hence
\be
\frac{\alpha_{s, \rm eff}(T, \mu)}{\alpha_{s, \rm eff}(T, 0)} > 1. \nonumber
\ee
The presence of a chemical potential might then invalidate perturbative calculations that work at zero  $\mu$. Taken literally, for each fixed value of the temperature there exists a (very high) density where the screening part of the quarks will eventually overcome the gluonic antiscreening and reverse asymptotic freedom.  
\pp
It is useful to recall at this stage the hard thermal/dense loop (HTL/HDL) perturbative results on the effective charge. From the static limit of the longitudinal part $\Pi_L(k^0, \vec k; T, \mu)$ of the gluon polarisation tensor one finds ($k = |\vec k |$) \cite{LeBellac:1996}
\be
\alpha_{s, \rm eff}(k^2; T, \mu) = \frac{\alpha_s}{\displaystyle 1 + \frac{\Pi_L(0, k ; T, \mu)}{k^2} } = \frac{\alpha_s}{\displaystyle 1 + \frac{m_D^2(T, \mu)}{k^2} } \label{alpha_s_eff_HTL}
\ee
with the Debye screening mass
\be
m_D^2(T, \mu) = \left(\frac{N_c}{3} +  \frac{N_f}{6} \right) g^2 T^2  + \frac{N_f}{2 \pi^2} g^2 \mu^2. \label{m_D}
\ee
Note that eq.(\ref{alpha_s_eff_HTL}) is both momentum- and medium-dependent. Since gluon and quark contributions to $m_D^2$ have the same sign, both contributions screen,  and the ratio 
\be
\frac{\alpha_{s, \rm eff}^{\rm HTL}(k^2; T, \mu)}{\alpha_{s, \rm eff}^{\rm HTL}(k^2; T, 0)} < 1 \quad \mbox{for all } k^2. \nonumber
\ee
Correspondingly, nothing spectacular happens when $T$ and/or $\mu$ are taken very large.
\pp
Clearly, various physical arguments suggest different and partly contradictory statements about the nature of $\alpha_s(T, \mu)$, which justifies in itself a detailed analysis. Also, in view of the importance of $\alpha_s(T, \mu)$ for future calculations, it is desirable to have a more refined calculation that aids to set quantitatively the in-medium scale $\Lambda_*^f$. In this paper, we will extend the formalism of refs.\cite{Schneider:2002,Schneider:2003}, that has already proven to work very successfully at $\mu = 0$, to finite chemical potential. A great advantage of the method lies in the fact that the extension is unique and physically intuitive. Section 2 briefly reviews the basics of the calculation. In section 3, we treat the high density limit at $T=0$, section 4 finally features the calculation of the physically relevant case of high temperature and $ \mu \lta T$. 
%
%
%
\section{Basics}
%
%
%
In \cite{Schneider:2002,Schneider:2003}, we have extended the approach of refs.\cite{Nielsen:1981,Petersen:1998} to finite temperature. Instead of a loop expansion of the gluon self-energy, the thermal energy shift
\be
\Delta E(H, T) = - \frac{1}{2} \left[ 4\pi \chi(H, T) \right] VH^2 - E(T)_{\rm vac}, \label{master1}
\ee
 of the perturbed thermal vacuum to order $\alpha_s$, after applying an external
static chromomagnetic field $H$, was evaluated at temperatures $T \gg \Lambda_{\rm QCD}$. The calculation of $\Delta E$ involves a sum over all Landau levels
\be
\bar \omega_{n,k_3, s_3} = \sqrt{k_3^2 + 2eH \left( n + 1/2 + s_3 \right)}, \label{E_quark}
\ee
weighted by the corresponding thermal occupation probabilities. Here, $n$ labels the mode, $k_3$ is the 3-component of the momentum and $s_3$ is the $z$-component of the spin of the particle under consideration.  Identifying $2eH$ with the scale $k^2$ at which the system is probed, as at $T=0$ ($e$ is the product of the strong coupling constant $g$ times the charge number $q$ that involves the structure constants of SU($N_c$)), we have investigated the high temperature limit
\be
\frac{eH}{T^2} \ll 1. \label{high_T}
\ee
Note that condition (\ref{high_T}) is equivalent to the hard thermal loop (HTL) approximation in thermal perturbation theory. Subsequently, we have extracted a temperature- and momentum-dependent dielectric permittivity $\epsilon(\vk, T)$ by use of the relation
\be
\alpha_{s, \rm eff}(\vk, T) \equiv \frac{\alpha_s}{\epsilon(\vk, T)} = \frac{\alpha_s}{1 - 4\pi \chi(\vk,
T)}, \label{eps_mu_T}
\ee
which is also valid at zero temperature. Here, $\chi(\vk, T) = \chi_g + \chi_q$ is the sum of gluon and quark magnetic susceptibilities. An expression similar to (\ref{master1}), the pure quantum part, persists even at zero temperature and has to be added to (\ref{master1}) in order to obtain the total result, so  $\chi_{\rm total} = \chi_0(k, \Lambda) + \chi(k, T)$. The  zero temperature susceptibility $\chi_0(k, \Lambda)$ has been calculated in \cite{Nielsen:1981,Petersen:1998} and results, using (\ref{eps_mu_T}), in the famous QCD running coupling expression:
\be
4 \pi \chi_0(H, \Lambda) = 4 \pi \left[ \chi_0^g(H, \Lambda) +  \chi_0^q(H, \Lambda) \right] = - \frac{g^2}{48 \pi^2} \left[ 11 N_c \log \left( \frac{2eH}{\Lambda^2} \right) - 2N_f \log \left( \frac{2eH}{\Lambda^2} \right) \right].  \label{chi_0}
\ee
$\Lambda$ corresponds to a zero temperature scale that characterises the ``medium'' and is ultimately identified with the renormalisation reference point of the coupling. We will be concerned with the in-medium part of the quarks, $\chi_q(k, T, \mu),$ in the following and add the vacuum contribution $\chi_0$ and the gluon contribution $\chi_g(k, T)$ only in the very end.

%

%
%
%
\section{The high density limit}
%
%
%
We will first evaluate the high density limit, i.e. $T = 0$ and
$$
\frac{eH}{\mu^2} \ll 1.
$$
For simplicity, we assume isospin symmetry in the following. 
The quark chemical potential acts only on the quark energy density, therefore we sum up all energy levels up to the Fermi surface:
\begin{eqnarray}
\Delta E(H, \mu)  & = & E^f - E^f_0, \ \nonumber \\
E^f & = & \sum_{n,k_2,k_3,s_3} \bar \omega_{n, k_3, s_3} \ \left[ \theta(\bar \omega - \mu) + \theta( \bar \omega + \mu)  \right]  \quad \mbox{and}  \nonumber \\
E_0^f & = & \sum_{k_1,k_2,k_3,s_3} \omega_k \ \left[\theta(\omega_k - \mu) + \theta(\omega_k + \mu) \right], \nonumber
\end{eqnarray}
with $\theta(x)$ the Heaviside step function, and summation over flavours is implicit. Apparently, antiparticles do not contribute to $\Delta E$. For the unperturbed case, $\omega_k = | \vec k | \equiv k$, and
\be
E_0^f = V \left[ 2 N_f \int \frac{d^3 k }{(2\pi)^3} |k| \theta(k - \mu) \right] =  V \frac{N_f}{4\pi^2} \mu^4. \nonumber
\ee
This expression has to be subtracted from 
\be
E^f  = \frac{V(eH)}{4\pi^2} \sum \limits_{s_3 = \pm 1/2} \ \sum \limits_{n=0}^\infty \int \limits_{- \infty}^{\infty} dk_3 \ \bar \omega_{n, k_3, s_3} \ \theta(\bar \omega - \mu), \nonumber
\ee
where we have replaced the summation over the quantum numbers $k_2$ and $k_3$ by an integral and a corresponding density of states, as outlined in \cite{Schneider:2002}. Performing the spin sum, separating off the $n=0$ contribution and transforming to dimensionless variables, we find
\be
E^f = V(eH) \left[\frac{\mu^2}{4\pi^2}  + \frac{2eH}{\pi^2} \sum \limits_{n=0}^\infty \int \limits_0^\infty dx \ \sqrt{x^2 + n + 1} \ \theta(\sqrt{x^2 + n +1} - \tilde \mu) \right], \quad \tilde \mu = \frac{\mu}{\sqrt{2eH}}. \nonumber
\ee
This expression is still exact. To proceed, we have to make some approximations. As in \cite{Schneider:2002}, we first simply replace the sum over $n$ by an integral. Again, this step will turn out to be insufficient to capture all effects in $e^2$, and we will give the complete result below, with corrections to the replacement taken into account. For now, we obtain after transforming to polar coordinates, 
\be
E^f = V(eH) \left[\frac{\mu^2}{4\pi^2}  + \frac{4eH}{\pi^2}  \int \limits_0^\infty dr \ \sqrt{r^2 + 1} \ \theta(\sqrt{r^2 +1} - \tilde \mu) \right]. \nonumber
\ee
Since we work in the high density limit, $\tilde \mu \gg 1$, the integral is dominated by the large $r$ limit. Approximating $\sqrt{r^2 +1} \sim r$, we find
\be
E^f = V(eH) \frac{\mu^2}{4\pi^2}  + V \frac{\mu^4}{4\pi^2}. \nonumber
\ee
The term $\propto \mu^4$ is just $E_0^f$ that has to be subtracted -- an important check of the consistency of the result. Summing over flavours and charges, the final result reads
\be
4\pi \chi(H, \mu) = - \frac{g^2 N_f}{2eH} \frac{\mu^2}{2 \pi^2} = - \frac{m_D^2(0, \mu)}{2eH}\nonumber
\ee
with the Debye screening mass (\ref{m_D}) at zero temperature and high density. With the help of eq.(\ref{eps_mu_T}), we recover the perturbative result (\ref{alpha_s_eff_HTL}), exactly as in \cite{Schneider:2002}, when we neglect corrections to the density of states and the higher-lying Landau levels. 
\pp
We now take into account these corrections: first, we use the Euler-MacLaurin formula
\be
\sum \limits_{n=0}^\infty f(n) = \int \limits_0^\infty f(n) dn + \frac{1}{2}\left[f(N) + f(0) \right] + \frac{1}{12} \left[ f'(N) - f'(0) \right] + ... \label{Euler}
\ee
to get the density of states correctly up to order $e^2$:
\bea
E^f & = & V(eH) \frac{\mu^2}{4\pi^2}  + V(eH)^2  \frac{2}{\pi^2}  \left[ 2 \int \limits_1^{\tilde \mu} du \ u^2 \sqrt{u^2 - 1} + \frac{1}{2} \int \limits_1^{\tilde \mu} du \ \frac{u^2}{\sqrt{u^2 -1}}  - \right. \nonumber \\ & & - \left. \frac{1}{24} \int \limits_1^\infty  du \frac{1}{\sqrt{u^2 -1}} \left\{\theta(u - \tilde \mu) + u \delta(u - \tilde \mu)  \right\} \right].\nonumber
\eea
Second, we evaluate the integrals, expand the results in the small parameter $1/\tilde \mu^2 = \mathcal{O}(e)$ and keep all terms up to order $e^2$. Then, 
\be
E^f  =  V(eH) \frac{\mu^2}{4\pi^2}  + V(eH)^2  \frac{2}{\pi^2}  \left[ \frac{\mu^4}{8(eH)^2} - \frac{\mu^2}{8eH} - \frac{5}{48} - \frac{1}{48} \log \left(4 \frac{\mu^2}{2eH} \right)  \right].\nonumber
\ee
Despite the very different structure of the integrals, the final result bears a surprising resemblence to the calculation at high temperature and zero density: All terms of order $\mu^2$ cancel, and the remainder shows the characteristic log structure. Hence, 
\be
4 \pi \chi_q(H, \mu) = g^2 \frac{2 N_f}{48 \pi^2} \log \left( \frac{[\mathcal{B}^f \mu]^2}{2eH} \right), \quad \mathcal{B}^f = 2 e^{5/2} \simeq 24.4
\ee
As in the high temperature calculation, the pre-factor of the log is exactly the same as in the zero-temperature case, so, including $\chi_0(H, \Lambda)$ of eq.(\ref{chi_0}) and trading $2eH$ for the momentum scale $k^2$, the final result reads
\be
\alpha_{s}^{\rm eff}(k^2, \Lambda; \mu) = \frac{\alpha_s(\Lambda)}{1 + {\displaystyle \frac{\alpha_s(\Lambda)}{12\pi} \left[ 11 N_c \log \left(\frac{k^2}{\Lambda^2} \right) - 2 N_f  \log \left(\frac{[\mathcal{B}^f \mu]^2}{\Lambda^2} \right) \right] }}, \label{alpha_s_mu}
\ee
What is somewhat surprising is the large number $\mathcal{B}^f$, which we expected to be $\mathcal{O}(1)$ from eq.(\ref{scale_1}). Note that the final result (\ref{alpha_s_mu}) is both $k^2$- and $\mu^2$-dependent, so it defies a simple interpretation like ``take the zero temperature, zero density running coupling $\alpha_s(k^2)$ and simply replace $k^2$ by $\mu^2$''. The situation at finite density is different from the high temperature case, where both gluons and quarks acquire a thermal average momentum, since it singles out the quarks. Furthermore, the large number $\mathcal{B}^f$ suggests that the antiscreening of the gluons will be  largely compensated due to the very large average momentum of the quarks. It is therefore not obvious when $\alpha_s(k^2, \mu)$ will become so small such as to justify perturbative calculations. 
%
%
%
%
\section{High temperature and moderate density}
%
%
Next, we explore the combined effects of temperature and chemical potential. The finite temperature gluon contribution 
\be
4\pi \chi_g(H, T) = - g^2 \frac{11 N_c}{48 \pi^2} \log \left( \frac{ [\Lambda_*^g]^2 }{2eH} \right) \label{chi_g}
\ee
has been dealt with in \cite{Schneider:2002,Schneider:2003} and is unaffected to this order by $\mu$, so we investigate the quark sector only. Again, we start from
\begin{eqnarray}
\Delta E(H, T, \mu) & = & E^f - E^f_0, \nonumber\\
E^f & = & \sum_{n,k_2,k_3,s_3} \bar \omega_{n, k_3, s_3} \ \left[ f_D(\bar \omega - \mu) + f_D( \bar \omega + \mu)  \right]  \quad \mbox{and} \nonumber  \\
E_0^f & = & \sum_{k_1,k_2,k_3,s_3} \omega_k \ \left[f_D(\omega_k - \mu) + f_D(\omega_k + \mu) \right]. \nonumber
\end{eqnarray}
The unperturbed energy is simply that of the non-interacting fermion gas,
\be
E_0^f(T, \mu) = V T^4 N_f \left[ \frac{7 \pi^2}{60} + \frac{1}{2} \frac{\mu^2}{T^2} + \frac{1}{4 \pi^2} \frac{\mu^4}{T^4}  \right],
\ee
which is easily obtained from eq.(\ref{F_5}), since
$$
E_0^f = V T^4  N_f \left[ \frac{24}{\pi^2} \lim_{\epsilon \rightarrow 0} F_5(\epsilon, (\mu/T)/\epsilon) \right].
$$
Now, in appropriate dimensionless units,
$$
E^f(H, T, \mu) = V T^4 \frac{\epsilon^2}{4 \pi^2} \sum \limits_{s_3 = \pm 1/2} \sum \limits_{n=0}^\infty \int \limits_0^\infty dx \sqrt{x^2 + \epsilon^2(n + 1/2 + s_3)} \ f_D^\pm(\sqrt{x^2 + \epsilon^2(n + 1/2 + s_3)}), \label{E_f_T_mu}
$$
where
$$
\epsilon^2 = \frac{2eH}{T^2} \quad \mbox{and} \quad \bar \mu = \frac{\mu}{T},
$$
and we defined, for the sake of lucidity,
\be
f_D^\pm(x) = \frac{1}{\exp(x - \bar \mu) + 1} +  \frac{1}{\exp(x + \bar \mu) + 1}.
\ee
After performing the spin sum, 
\be
E^f = V T^4 \left( \frac{\epsilon^2}{24} + \frac{\epsilon^2  \bar \mu^2}{8 \pi^2} +  \frac{\epsilon^2}{2 \pi^2} \sum \limits_{n=0}^\infty \int \limits_0^\infty dx \sqrt{x^2 + \epsilon^2(n + 1)} \ f_D^\pm(\sqrt{x^2 + \epsilon^2(n + 1)}) \right), \label{E_f_T_mu_2}
\ee
which is still exact. In the following, we explore the high temperature, moderate density limit, i.e. 
\be
\epsilon^2 \ll 1 \quad \mbox{ and } \bar \mu \lta 1,
\ee
and calculate an approximation to eq.(\ref{E_f_T_mu_2}) up to order $\epsilon^4 \bar \mu^2 = g^2 (\mu/T)^2$. 
\pp
We proceed in close analogy to the high density case: first, we simply replace the sum by an integral. In our HTL/HDL approximation, we neglect corrections to the sum and the Landau levels to arrrive, after summing over the couplings, at
\be
E_f = - \frac{1}{2} V H^2 \left\{ - \frac{N_f}{6} \frac{g^2 T^2}{2eH}  - \frac{N_f}{2 \pi^2} \frac{g^2 \mu^2}{2eH} \right\} + E_0^f.
\ee
The quantity in curly brackets is already $4 \pi \chi_q(H, T, \mu)$, which leads, when including the gluon susceptibility $4 \pi \chi_g(H, T)$ in the corresponding approximation \cite{Schneider:2002} and setting $k^2 = 2eH$,  to an effective coupling
\be
\alpha_{s, \rm eff}(k^2; T, \mu) = \frac{\alpha_s}{\displaystyle 1 + \frac{m_D^2(T, \mu)}{k^2} }
\ee
with the HTL/HDL Debye mass (\ref{m_D}). Recovering again the perturbative result by exactly the same procedure as at high temperature or high density, respectively, is an important check of the internal consistency of the used method here. Our incorporation of the chemical potential in the calculation is therefore well founded, which adds credibility to the complete result of $\mathcal{O}(g^2)$, to which we turn now. 
\pp
Following our previous approach, we first take into account corrections to the density of states according to eq.(\ref{Euler}):
\bea
E^f & = & V T^4 \left( \frac{\epsilon^2}{24} + \frac{\epsilon^2  \bar \mu^2}{8 \pi^2} + \frac{1}{\pi^2} \left[ 24 F_5(\epsilon, \bar \mu / \epsilon ) + 2 \epsilon^2 F_3(\epsilon, \bar \mu / \epsilon) \right] + \right. \nonumber \\ & & + \left. \frac{\epsilon^2}{4 \pi^2} \left[ 2 F_3(\epsilon, \bar \mu / \epsilon ) + \epsilon^2 F_1(\epsilon, \bar \mu / \epsilon) \right] - \frac{\epsilon^4}{48 \pi^2} H(\epsilon, \bar \mu / \epsilon) \right), \label{full_E_f}   
\eea
using the expressions for the $F_i$ and $H$, as defined in the appendix. Second, we now expand the subsequent integrals in the small parameter $\epsilon$ with the help of eqs.(\ref{F_1}), (\ref{F_3}), (\ref{F_5}) and (\ref{H}) up to order $\epsilon^4$  to arrive at the simple expression
\be
\Delta E^f(H, T, \mu) = \Delta E^f(H, T, \mu = 0) + V T^4 \left( - \frac{7 \zeta(3)}{192 \pi^4} \epsilon^4 \bar \mu^2 \right), 
\ee
with $\zeta(3) \simeq 1.202...$ All terms of order $\epsilon^2 \bar \mu^2$ cancel identically. Taking into account $\chi^q_0(H, \Lambda)$ of eq.(\ref{chi_0}), we obtain the following form for the total quark susceptibility:
\be
4 \pi \chi_q(T, \mu) = g^2 \frac{2 N_f}{48 \pi^2} \left[ \log \left( \frac{[\bar \mathcal{A}_f \pi T]^2}{\Lambda^2}  \right)  + \frac{7 \zeta(3)}{2 \pi^2} \frac{\mu^2}{T^2} \right],\label{final}
\ee
which is $H$-independent, as in the high temperature, zero density case. This is a highly non-trivial result that only arises because of intricate cancellations of terms other than $\mathcal{O}(\epsilon^4)$. The persistence of scale-independence even in the presence of a chemical potential is a remarkable outcome.  
%
\section{Conclusions}
%
%
We have presented for the first time a self-contained calculation of a medium-dependent QCD running coupling constant $\alpha_s(T, \mu)$ at finite quark chemical potential, based on the semi-classical background field method of refs.\cite{Schneider:2002,Schneider:2003}. We have considered the two cases high density, zero temperature (i.e. $\mu^2/H \gg 1$ and $T=0$) and high temperature, moderate density (i.e. $T^2/H \gg 1$ and $\mu/T \lta 1$), and have calculated the induced energy shift $\Delta E^f$ of the vacuum in the quark sector. In both cases, the corresponding HTL/HDL perturbative results were recovered by neglecting corrections to the density of states and the higher-lying Landau levels, as in the high temperature, zero density case. The approach is therefore internally consistent, and the inclusion of the quark chemical potential is free from ambiguities and sensible. 
\pp
Taking into account the mentioned corrections, all $H$- and hence scale-dependence drops out of the in-medium contributions once the vacuum (zero temperature, zero density) contributions are added. In addition, the running coupling becomes very simple in both cases, as at high temperature, zero density: it again follows from the zero temperature renormalisation group equations, with the momentum scales of the propagating particles replaced by appropriate in-medium scales $\Lambda_*$. At high density, the gluon part remains unmodified, whereas the quark part has the form 
$$
- g^2 \frac{2 N_f}{48 \pi^2}  \log \left(\frac{[\Lambda_*^f]^2}{\Lambda^2} \right),
$$
with the in-medium scale $\Lambda_*^f = 2 e^{5/2} \mu \simeq 24.4 \ \mu$. 
\pp
At high temperature and moderate density, the final result that follows from (\ref{final}), including the gluon contribution $\chi_0^g + \chi_g(T)$, can be written as
\be
\alpha_{s}^{\rm eff} = \frac{\alpha_s}{1 + {\displaystyle \frac{\alpha_s}{12\pi} \left[ 11 N_c \log \left(\frac{[\Lambda_*^g]^2}{\Lambda^2} \right) - 2 N_f  \log \left(\frac{[\Lambda_*^f]^2}{\Lambda^2} \right) \right] }}. \label{alpha_s_T_mu_final}
\ee
Here, the gluon in-medium scale is obviously the same as in the high temperature, zero density case, i.e.
$$
\Lambda_*^g \simeq 0.938 (2 \pi T).
$$
The quark in-medium scale receives a correction due to the chemical potential:
\be
\left[ \Lambda_*^f \right]^2 = \exp\left( \bar \mathcal{B}_f \frac{\mu^2}{T^2} \right)  \left(\bar \mathcal{A}_f \pi T \right)^2, \quad \bar \mathcal{B}_f = \frac{7 \zeta(3)}{2 \pi^2} \simeq 0.43,
\ee
which can, for $\mu/T \lta 1$, be expanded as
\be
\left[\Lambda_*^f \right]^2 \simeq [2.91 \ T]^2 + [1.91 \ \mu]^2. \label{Lambda_f_mu}
\ee
This form of the quark in-medium scale $\Lambda_*^f$ is surprisingly reminiscent of our initial estimate (\ref{scale_1}). We have therefore been able for the first time to give a sound (though semiclassical) justification for approximate use of the naive phenomenological quark scale $\Lambda^2_* = (\pi T)^2 + \mu^2$ in the zero temperature, zero density running coupling, although in our calculation the influence of the chemical potential is somewhat larger (which is, however, in accord with the result at finite density and zero temperature, see above). 
\pp
To get a quantitative feeling, we show in figure 1 the effective coupling $g_s^{\rm eff}(T, \mu)$ as a function of $T / T_c$, for $\mu = 500$ MeV. There, $g_s(T, \mu)$ (solid line) is our exact result (\ref{alpha_s_T_mu_final}), $g_s(T, 0)$ the same at $\mu =0$ (dashed line). Apparently, the coupling strength increases, as anticipated, at finite chemical potential due to the larger average in-medium momentum of the quarks, and the effect becomes most pronounced at small temperatures. For comparison, $g_s^{\rm phen}(T, \mu)$ corresponds to the phenomenological expectation, i.e. $\Lambda_*^g = 2\pi T$ and $\Lambda_*^f = \sqrt{(\pi T)^2 + \mu^2}$ (dash-dotted line). Applying the full result (\ref{alpha_s_T_mu_final}) still leads to a coupling strength $g_s$ larger than one in the temperature region where lattice results on the equation of state at finite $\mu$ exist. It might nevertheless serve as quantitative input for actual perturbative calculations that will try to describe these data. 
\begin{figure}
\begin{center}
\epsfig{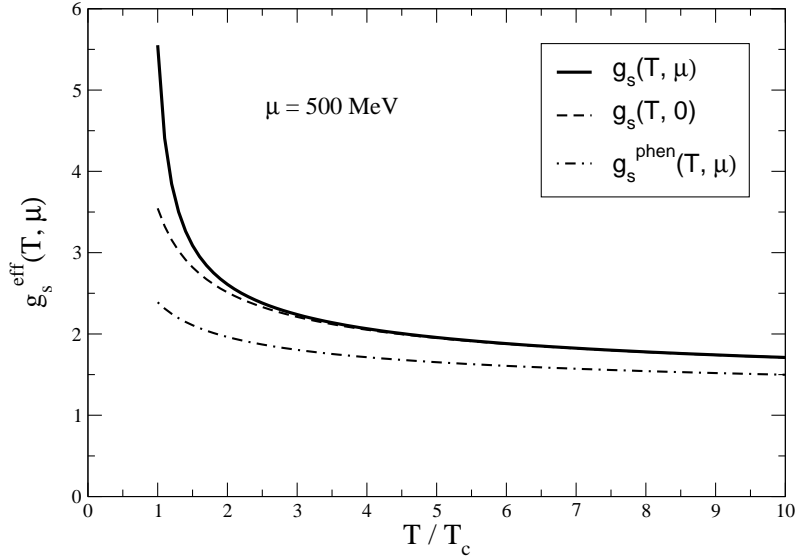}
\caption{The effective coupling $g_s^{\rm eff}(T, \mu)$ as a function of $T / T_c$. $T_c$ is taken to be 170 MeV, $\Lambda_{\rm QCD} \simeq 250$ MeV and $\mu = 500$ MeV. For explanations, see text.}
\end{center}
\end{figure}
%

%
%
\section{Appendix}
%
%
To make this paper self-contained, we present here the formulas used to evaluate the $\epsilon$- and $\bar \mu$-dependent integrals, appearing, e.g., in eq.(\ref{full_E_f}). Since $\epsilon \ll 1$ plays the role of a small mass term, the whole machinery of the small $m/T$ expansion of thermodynamical integrals can be applied. Our approach correspondingly simply extends the calculations of \cite{Haber:1981}, where the expansions were carried out for bosonic integrals, to the fermionic case. We will be short on some details, hence, but comment on differences where appropriate. 
We start from 
\be
g_n(y, r) \equiv \frac{1}{\Gamma(n)} \int \limits_0^\infty dx \ x^{n-1} \left[ \frac{1}{\exp(\sqrt{x^2 + y^2} - ry) + 1} \right] \label{g_n_def}
\ee
and
\be
f_n(y, r) \equiv \frac{1}{\Gamma(n)} \int \limits_0^\infty dx \frac{x^{n-1}}{\sqrt{x^2 + y^2}} \left[ \frac{1}{\exp(\sqrt{x^2 + y^2} - ry) + 1} \right]. \label{f_n_def}
\ee
Here, $\Gamma(n)$ is the Gamma function. It is useful to define
\be
G_n(y, r) = g_n(y, r) - g_n(y, -r) \label{G_n_def}
\ee
and
\be
F_n(y, r) = f_n(y, r) + f_n(y, -r). \label{F_n_def}
\ee
These integrals appear in thermodynamical expressions for an ideal gas of fermions with mass $m$ at finite chemical potential. The corresponding dimensionless quantities are therefore $y = m/T$ and $r = \mu/m$. In the following, we will be concerned with the high-temperature limit of eq.(\ref{F_n_def}), i.e. $m/T \ll 1$ and $\mu \lta T$, for odd $n$. In the bosonic case, the physical region is $| r | \leq 1$, otherwise Bose-Einstein condensation phenomena have to be taken into account. For fermions, no such restriction exists, so $r$ can be in principle arbitrarily large. The expansion of (\ref{G_n_def}) (and, correspondingly, (\ref{F_n_def})) in small $y$ will have the general form
\be
I_n(y,r) = \tilde I_n(y,r) + \sum \limits_{k = 0}^\infty a_k(r) \left( \frac{y}{\pi} \right)^k \quad (I = F, G), \label{I_n_general}
\ee
where $\tilde I_n$ might be a non-polynomial function in $y$, and the $a_k(r)$ are themselves polynomials in $r$. If $a_k(r)$ is of the form
$$
a_k(r) = \sum \limits_{i=0}^k b_i r^i, \quad \mbox{with } b_k \neq 0,
$$
expression (\ref{I_n_general}) also becomes an expansion in $ry = \bar \mu = \mu/T$. To be able to capture all leading effects in $\mu$, we will therefore restrict ourselves in the following to the situation $\bar \mu \lta 1$.
\pp
$G_n$ and $F_n$ obey the same recursion relations as their bosonic counterparts:
\bea
\frac{d}{dy} G_{n+1} & = & r n F_{n+1} + \frac{y^2 r}{n} F_{n-1} - \frac{y}{n} G_{n-1},  \label{recurs_G} \\
\frac{d}{dy} F_{n+1} & = & \frac{r}{n} G_{n-1} - \frac{y}{n} F_{n-1}. \label{recurs_F}
\eea
With the initial conditions 
\be
G_n(0, r) = 0 \quad (n > 0), \quad F_n(0, r) = \frac{2(1 - 2^{2-n})}{n-1} \zeta(n-1) \quad (n > 2), \label{initial}
\ee
$\zeta(n)$ being the Riemann Zeta function, all necessary functions can therefore be derived from $G_1$ and $F_1$.  
In close analogy to \cite{Haber:1981}, we substitute the identity
\be
\frac{1}{e^z + 1} = \frac{1}{2} - \sum \limits_{n = - \infty}^{ \infty} \frac{z}{z^2 + (2n+1)^2 \pi^2} \label{FD_identity}
\ee
in the integrands of $G_1$ and $F_1$ and integrate term by term. We start with $G_1$:
\be
G_1(y,r ) = - 2 \sum \limits_{m = 0}^{\infty} \mathcal{G}_m(y,r),
\ee
where
\be
\mathcal{G}_m(y,r) = 2 y r \int \limits_0^\infty dx \frac{x^2 + y^2(1-r^2) - \bar \omega_m^2}{[ x^2 + y^2 (1-r^2) + \bar \omega_m^2]^2 + \bar \omega_m^2 (2yr)^2}  \label{G_m}
\ee
and the (dimensionless) Matsubara frequencies $\bar \omega_m = (2m+1)\pi$. Note that $\mathcal{G}_m$ has exactly the same functional form in the bosonic case, only there $\bar \omega_m = 2m \pi$. Next, the integrand in (\ref{G_m}) is expanded in powers of $y$ and integrated term by term. After that, the summation over $m$ can be performed, yielding $\zeta$ functions. The final result reads
\be
G_1(y, r) = - ry + 2 \pi r \sum \limits_{k=1}^\infty (-1)^k \mathcal{A}_k [1 - 2^{-(2k+1)}] \zeta(2k+1) \left( \frac{y}{\pi} \right)^{2k+1}, \label{G_1_full}
\ee
where $\mathcal{A}_1 = 1$, $\mathcal{A}_2 = 2r^2 + \frac{3}{2}$, $\mathcal{A}_3 = 3r^4 + \frac{15}{2} r^2 + \frac{15}{8}$. These coefficients are the same as in the bosonic case. 
\pp
For $F_1$, we need to introduce a convergence factor $x^{-\epsilon}$ ($0 < \epsilon < 1$) in the integrand. In the end, the result is finite, and the limit $\epsilon \rightarrow 0$ can be safely taken. We start from
\be
F_1(y,r) = \mathcal{F}(y,r) - 2 \sum \limits_{m=0}^{\infty} \mathcal{F}_m(y,r).
\ee
Now,
\be
\mathcal{F}(y,r) = \int \limits_0^\infty dx \frac{x^{- \epsilon}}{\sqrt{x^2 + y^2}} = \frac{1}{\epsilon} - \log \left( \frac{y}{2} \right)  + \mathcal{O}(\epsilon)
`\ee
and
\be
\mathcal{F}_m(y,r) = 2  \int \limits_0^\infty dx \ x^{- \epsilon} \frac{x^2 + y^2(1-r^2) + \bar \omega_m^2}{[ x^2 + y^2 (1-r^2) + \bar \omega_m^2]^2 + \bar \omega_m^2 (2yr)^2}.  \label{F_m}
\ee
The leading term in $y$ is
\be
\mathcal{F}_m^{(1)}(y,r) = 2 \int \limits_0^\infty dx \frac{x^{- \epsilon}}{x^2 + \bar \omega_m^2} = \frac{1}{\bar \omega_m^{1 + \epsilon}} \frac{\pi}{\cos(\pi \epsilon/2)}.
\ee
Using $\zeta(1 + \epsilon) = 1 / \epsilon + \gamma + \mathcal{O}(\epsilon)$, $\gamma = 0.5772...$ being the Euler-Mascheroni constant, the summation over the Matsubara frequencies yields
\be
- 2 \sum \limits_{m=0}^{\infty} \mathcal{F}_m^{(1)}(y,r) = - \frac{1}{\epsilon} - \gamma + \log\left( \frac{\pi}{2} \right), %
\ee
which removes the $1/\epsilon$ singularity of $\mathcal{F}(y,r)$. Higher order terms in $y$ are straightforward, and the final result reads
\be
F_1(y, r) = - \log \left( \frac{y}{\pi} \right) - \gamma - 2 \sum \limits_{k=1}^\infty (-1)^k \mathcal{B}_k [1 - 2^{-(2k+1)}] \zeta(2k+1) \left( \frac{y}{\pi} \right)^{2k}
\ee
with the universal coefficients $\mathcal{B}_1 = r^2 + \frac{1}{2}$, $\mathcal{B}_2 = r^4 + 3r^2 + \frac{3}{8}$, $\mathcal{B}_3 = r^6 + \frac{15}{2}r^4 + \frac{45}{8}r^2 + \frac{5}{16}$. Here, $\mathcal{B}_k \propto r^{2k}$, therefore the first term in $\bar \mu$ in $F_1$ starts at order $\bar \mu^2 y^0$. 
\pp
With eqs.(\ref{recurs_G}), (\ref{recurs_F}) and (\ref{initial}), all other terms of interest can be worked out to any order in $y$. We quote the expansions that are of interest in this work:
\be
F_1(y, r) = - \log \left( \frac{y}{\pi} \right) - \gamma + \frac{7 \zeta(3)}{4 \pi^2} \left[r^2 + \frac{1}{2} \right] y^2, \label{F_1}
\ee

\be
F_3(y,r) = \frac{\pi^2}{12} + \frac{y^2}{4} \left[ r^2 + \gamma - \frac{1}{2} + \log\left( \frac{y}{\pi} \right) \right] -  \frac{7 \zeta(3)}{16 \pi^2} \left[ r^2 + \frac{1}{4} \right] y^4, \label{F_3}
\ee

\be
F_5(y,r) = \frac{1}{12} \left( \frac{7\pi^4}{120} + \frac{ \pi^2}{4} \left[ r^2 - \frac{1}{2} \right] y^2 + \frac{3}{16} \left[ \frac{2}{3} r^4 - 2 r^2 - \gamma + \frac{3}{4} - \log \left(\frac{y}{\pi} \right)  \right] y^4 +  \frac{7 \zeta(3)}{256 \pi^2} \left[ r^2 + \frac{1}{6} \right] y^6 \right). \label{F_5}
\ee
Note that a term like $r^2 y^6$ is actually of order $(\mu/T)^2 (m/T)^4$, so it is necessary to push the expansion to higher terms in $y$ to capture all effects in $m/T$ to a desired order.
\pp
As in \cite{Schneider:2002}, for the evaluation of the derivative terms in eq.(\ref{full_E_f}) we also need the small $y$ expansion of integrals such as
\bea
H(y, r) & = &  \int \limits_0^\infty dx \left[ \left( \frac{1}{\sqrt{x^2 + y^2}} -  \frac{1}{1 + \exp( - [\sqrt{x^2 + y^2} - ry])} \right) \right. \\ & & \left. \times \frac{1}{\exp(\sqrt{x^2 + y^2} -ry) + 1} \right] + ( r \rightarrow -r) \label{formula_deriv} \\
& = & F_1(y,r) - \left[ \int \limits_0^\infty dx  \frac{\exp(\sqrt{x^2 + y^2} -ry)}{\left( \exp(\sqrt{x^2 + y^2} -ry) + 1 \right)^2} + ( r \rightarrow -r) \right].\label{formula_deriv2}
\eea
For the evaluation of the second term in (\ref{formula_deriv2}) we employ the same trick as in \cite{Schneider:2002}. Introduce a parameter $\alpha$ to write
\be
\bar H (y, r; \alpha) = - \left[ \int \limits_0^\infty dx \frac{\exp(\alpha \sqrt{x^2 + y^2} -ry)}{(\exp(\alpha \sqrt{x^2 + y^2} -ry) + 1)^2} + (r \rightarrow -r) \right]. \label{formula1}
\ee
Obviously, $\bar H(y, r; 1)$ is the sought quantity. Now we re-write
\bea
\bar H (y, r; \alpha) & = & \frac{\partial}{\partial \alpha} \left[ \int \limits_0^\infty dx \ \frac{1}{\sqrt{x^2 + y^2}} \ \frac{1}{\exp(\alpha \sqrt{x^2 + y^2} - ry) + 1} + (r \rightarrow -r) \right] \\ & = & \frac{d}{d \alpha} F_1(\alpha y, r/ \alpha) \\
& = & - \frac{1}{\alpha} + \alpha \frac{7 \zeta(3)}{4 \pi^2} y^2.
\eea
Therefore
\be
H(y, r) = - \log \left( \frac{y}{\pi} \right) - \gamma - 1 + \frac{7 \zeta(3)}{4 \pi^2} \left[r^2 + \frac{3}{2} \right]  y^2. \label{H}
\ee


\begin{thebibliography}{99}


\bibitem{Fodor:2002a} Z. Fodor, S.D. Katz and K.K. Szabo, hep-lat/0208078.

\bibitem{Fodor:2002b} Z. Fodor and S.D. Katz, JHEP 0203 (2002) 14.

\bibitem{Allton:2002} C.R. Allton et al., Phys. Rev. {\bf D66} (2002) 074507.

\bibitem{Peshier:1996} A.~Peshier, B.~K\"{a}mpfer, O.P. Pavlenko and G.~Soff, Phys.\ Rev.\  {\bf D54} (1996) 2399.

\bibitem{Levai:1998} P.~Levai and U.~Heinz, Phys.\ Rev.\ {\bf C57} (1998) 1879.

\bibitem{Schneider:2001}  R.A. Schneider and W. Weise, Phys. Rev. {\bf C64} (2001) 055201. 

\bibitem{Peshier:2002} A. Peshier, B. K\"ampfer and G. Soff, Phys. Rev. {\bf D66} (2002) 094003.

\bibitem{Letessier:2003} J. Letessier and J. Rafelski, hep-ph/0301099. 

\bibitem{Andersen:2001} J.O. Andersen, E. Braaten and M. Strickland, Phys. Rev. {\bf D63} (2001) 105008.

\bibitem{Blaizot:2001} J.P. Blaizot, E. Iancu and A. Rebhan, Phys. Rev. {\bf D63} (2001) 065003.

\bibitem{Schneider:2002} R.A. Schneider, Phys. Rev. {\bf D66} (2002) 036003.

\bibitem{Schneider:2003} R.A. Schneider, hep-ph/0210281, Phys. Rev. {\bf D} in print.

\bibitem{Nielsen:1981} N. K. Nielsen, Am. J. Phys. {\bf 49} (1981) 1171.

\bibitem{Petersen:1998} J. L. Petersen, in {\em Proceedings of the 1997 European Summer School of High-Energy Physics}
(Edts. N. Ellis and M. Neubert), CERN 98-03.

\bibitem{LeBellac:1996} M. Le Bellac, {\em Thermal Field Theory} (Cambridge University Press, 1996).

\bibitem{Haber:1981} H.E. Haber and H.A. Weldon, Phys. Rev. {\bf D25} (1982) 502. 

\end{thebibliography}
\end{document}